\documentclass{WileyMSP-template}
\newcommand{\cheers}{\textsc{cheers}}
\usepackage{amsmath,amssymb,mleftright,mathtools,microtype}
\usepackage[colorlinks,linkcolor=blue,citecolor=blue,urlcolor=blue]{hyperref}
\usepackage[dvipsnames,table]{xcolor}
\usepackage{graphicx}
\graphicspath{{./FigsRevision/}}
\usepackage{cleveref,chemformula,bm,booktabs}

\newcommand{\mathbfit}[1]{\bm{\mathit{#1}}}
\renewcommand{\vec}[1]{\mathbfit{#1}}
\newcommand{\Hh}{\hat{H}}
\newcommand{\bos}{\text{bos}}
\newcommand{\el}{\text{el}}
\newcommand{\elbos}{\text{el-bos}}
\newcommand{\hphi}{\hat{\phi}}
\newcommand{\hc}{\hat{c}}
\newcommand{\hd}{\hat{d}}
\newcommand{\ha}{\hat{a}}

\def\b{\beta}
\def\ex{{\rm e}}

\newcommand{\cG}{\mathcal{G}}
\newcommand{\mcG}{\bm{\mathcal{G}}}

\def\bra{\langle}
\def\ket{\rangle}
\def\bgm{\mbox{\boldmath $\mu$}}

\def\fbgm{\mbox{\scalebox{.7}{$\scriptscriptstyle \bgm$}}}

\begin{document}

\pagestyle{fancy}
\rhead{\includegraphics[width=2.5cm]{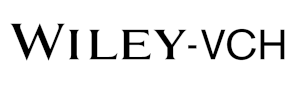}}

\title{Cheers: a linear-scaling KBE+GKBA code}

\maketitle

\author{Y. Pavlyukh}
\author{R. Tuovinen}
\author{E. Perfetto}
\author{G. Stefanucci}

\dedication{}

\begin{affiliations}
Y. Pavlyukh\\
Institute of Theoretical Physics, Faculty of Fundamental Problems of Technology, Wroclaw University of Science and Technology, 50-370 Wroclaw, Poland \\
yaroslav.pavlyukh@gmail.com\\
R. Tuovinen\\
Department of Physics, Nanoscience Center, P.O. Box 35, 40014 University of Jyv{\"a}skyl{\"a}, Finland\\
E. Perfetto and G. Stefanucci\\
Dipartimento di Fisica, Universit{\`a} di Roma Tor Vergata, Via della Ricerca Scientifica 1, 00133 Rome, Italy and
INFN, Sezione di Roma Tor Vergata, Via della Ricerca Scientifica 1, 00133 Rome, Italy
\end{affiliations}


\begin{abstract}
  The interaction of electrons with quantized phonons and photons underlies the ultrafast dynamics of systems ranging
  from molecules to solids, giving rise to a plethora of physical phenomena experimentally accessible using
  time-resolved techniques. Green's function methods offer an invaluable interpretation tool since scattering mechanisms
  of growing complexity can be selectively incorporated in the theory. \cheers{} is a general-purpose nonequilibrium
  Green's function code that implements virtually all known many-body approximations and is designed for first
  principles studies of ultrafast processes in molecular and model solid state systems. The aims of generality,
  extensibility, efficiency, and user friendliness of the code are achieved through the underlying theory development
  and the use of modern software design practices. Here, we motivate the necessity for the creation of such a code and
  overview its design and capabilities.
\end{abstract}

\section{Introduction}                            
Coherent electron dynamics in correlated materials is typically accompanied by the interaction with bosonic particles
and quasiparticles, such as phonons, plasmons, charge density waves, and photons. Numerous proposals have been pushed
forward for creating materials with novel properties by putting them into an entangled light-matter
state~\cite{basov_towards_2017}. Vastly different energy and length scales, quantum aspects of the involved bosonic
particles, and numerous intertwined orders pose considerable challenges for theory. A scalable theoretical method to
model excitation and relaxation phenomena in correlated many-body systems, reliable beyond the perturbative regime, is
therefore crucial to simulate and interpret experimental results and to suggest new materials to study.

Many-body perturbation theory represents a systematic way to deal with inter-particle correlations.  In order to get
access to the dynamical properties of the system, the equations of motion (EOM) for the two-times electron and boson
Green's functions (GF), hereafter referred to as the \emph{nonequilibrium Green's function} (NEGF)
theory~\cite{stefanucci_nonequilibrium_2013}, must be propagated. The EOMs in this case are known as the Kadanoff-Baym
equations (KBE).  The time non-locality of the scattering term represents the major difficulty for the full two-times
propagation leading to the scaling that is at least cubic ($t_{\text{f}}^3$) with the physical propagation time
$t_{\text{f}}$~\cite{kwong_real-time_2000,dahlen_solving_2007,myohanen_many-body_2008,galperin_linear_2008,
  myohanen_kadanoff-baym_2009,von_friesen_successes_2009,bittner_coupled_2018} making it very difficult to resolve
smaller energy scales associated, for instance, with phonons~\cite{schuler_time-dependent_2016}. Still, interesting
progress has recently been made~\cite{schuler_nessi_2020,kaye_low_2021,
  meirinhos_adaptive_2022,dong_excitations_2022}. The so-called generalized Kadanoff-Baym ansatz
(GKBA)~\cite{lipavsky_generalized_1986} allows one to limit the propagation to time-diagonal, that is, to work with
one-particle density matrices rather than with two-times Green's functions, preserving concomitantly the description of
inter-particle correlations. First-principles implementations of the KBE+GKBA equations have been pioneered in atoms and
organic molecules in Refs.~\cite{bostrom_charge_2018,covito_real-time_2018,latini_charge_2014,maansson_real-time_2021,
  perfetto_ultrafast_2018,perfetto_first-principles_2019, perfetto_ultrafast_2020,perfetto_first-principles_2015} and 2D
materials~\cite{perfetto_real-time_2022,perfetto_real-time_2023} using the \cheers{} (a tool for the correlated
hole-electron evolution from real-time simulations) code~\cite{perfetto_cheers:_2018}.

KBE+GKBA became a competitive first-principles method in pure electronic case when formulated in an ordinary
differential equation (ODE) form with linear time-scaling~\cite{schlunzen_achieving_2020,
  joost_g1-g2_2020,pavlyukh_photoinduced_2021}.  In our recent Letter~\cite{karlsson_fast_2021}, we generalized the GKBA
to quantized bosonic particles.  In Ref.~\cite{tuovinen_time-linear_2022}, we further extended the time-linear
formulation to quantum transport, empowering the Meir-Wingreen formula~\cite{meir_landauer_1992} for calculating
time-dependent currents. The second version of \cheers{} represents a novel implementation of these ideas adding the
time-linear formulation, the interaction with bosonic particles as well as the coupling to fermionic baths. The latest
improvement of theory~\cite{pavlyukh_time-linear_2022-1,pavlyukh_time-linear_2022,pavlyukh_photoinduced_2021} was a
natural reason for restructuring of the code and incorporation of new features such as computation of the molecular
integrals based on our earlier implementation~\cite{pavlyukh_electron_2013}. In what follows we will denote our
computational scheme as GKBA+ODE in order to emphasize the time-linear implementation.

It may be tempting to conclude that GKBA+ODE is a black box method that can be applied to a variety of systems without
any physical insight. One can even parallel it with another popular approach, time-dependent density functional theory
(TDDFT). Common to them are a small set of observables, linear propagation time, and reliance on additional
approximations. The knowledge of these approximations distinguishes expert user from a casual practitioner. They should
be carefully selected based on the level of correlations in the initial and excited state of the studied
system~\cite{pavlyukh_photoinduced_2021}, and one should be aware that GKBA, while maintaining the conserving properties
of the many-body approximations, may lead to violations of other physically mandated properties, such as the positivity
of electronic~\cite{joost_dynamically_2022} and doublonic~\cite{pavlyukh_time-linear_2022} occupation numbers.

The purpose of this contribution is to assist users in taking their first steps with the \cheers{} code and to
explain its structure to developers. Technical details of the underlying theory and implementation can be found in the
Supporting Information.

\section{Structure of the code}
Mathematically speaking, GKBA+ODE represents an initial value problem (ivp)
\begin{subequations}
  \label{eq:ivp}
\begin{align}
  \dot{\vec{y}}(t)&=\mathcal{F}\mleft\{t, \vec{y}(t)\mright\},\label{eq:ode}\\
  \vec{y}(t_0)&=\vec{y}_0,
\end{align}
\end{subequations}
where $\mathcal{F}$ depends on the time $t$ and is a complicated functional of the state vector $\vec{y}(t)$. Under the
state vector we understand a collection of electronic, bosonic and embedding correlators that fully characterize the
system under consideration. Its precise composition is not fixed and depends on the level of approximation.

The goal is to compute the evolution in time of some relevant physical observables depending on the system state, that
is
\begin{align}
  \vec{O}(t)&=\vec{O}\mleft\{\vec{y}(t)\mright\}.
\end{align}
Typical examples of observables would be the dipole transition moment, different contributions to the total energy,
electron and doublon occupation numbers, a set of bosonic observables such as occupation numbers and displacements, and
currents for open systems.

Under some mild conditions, the ivp has a unique solution. Thus, at least mathematically, the problem is well-posed and
requires a mere application of some well-known methods such as Runge-Kutta (RK) for the approximate solutions of
ordinary differential equations (ODE). While this indeed is the essential part of the \cheers{} code, the computational
complexity of the $\mathcal{F}$ functional grows rapidly with the system size. This is due to the fact that the state
vector $\vec{y}(t)$ contains the one- and two-body density matrices with dimensions growing as a second and forth power
of the system size, respectively. It represents the major challenge for implementation and determines the feasibility of
the method as well as the maximal propagation time $t_{\text{f}}$.  In what follows, we characterize the system with the
number of electronic $N_\text{$e$-basis}$ and bosonic $N_\text{$b$-basis}$ basis functions and denote the dimension of
the state vector $\vec{y}(t)$ as $N_v$. Already the form of Eqs.~\eqref{eq:ivp} and the dimension $N_v$ mandate important design
decisions.
\subsection{General design features}
\begin{table}
  \caption{\label{tab:aminoacids} Typical characteristics of organic molecules (amino
    acids) computed using the pVDZ basis set (for tyrosine 6311G is used). $n_o$ and $n_v$
    represents typical number of occupied and virtual states, respectively.}
    \arrayrulecolor{NavyBlue}
     \begin{tabular}{rcrccc}\toprule
       System& Structure& Formula&AO basis & $n_o$ & $n_v$\\\midrule
       Alanine &\raisebox{-0.3cm}{\includegraphics[width=1.3cm]{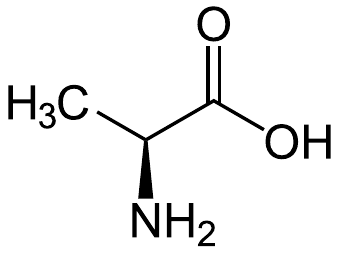}} &\ch{C3NH7O2} & 125 & 18 &17\\
       Glycine&\raisebox{-0.4cm}{\includegraphics[width=0.8cm]{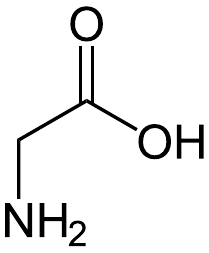}} &\ch{C2NH5O2} & 100 & 15 &15\\
       Tyrosine&\raisebox{-0.2cm}{\includegraphics[width=2.2cm]{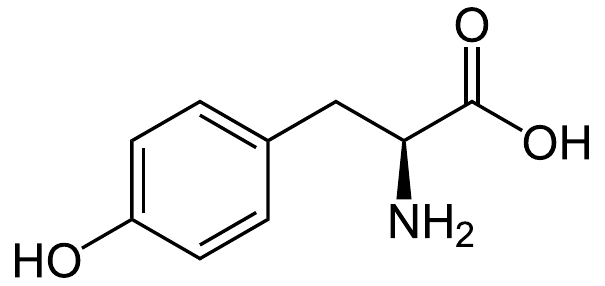}} &\ch{C9NH11O3} & 202 & 35 & 35\\
       Proline&\raisebox{-0.2cm}{\includegraphics[width=1.2cm]{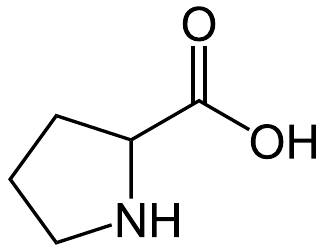}} &\ch{C5NH9O2} & 165 & 23 & 24\\
       Phenylalanine&\raisebox{-0.2cm}{\includegraphics[width=1.8cm]{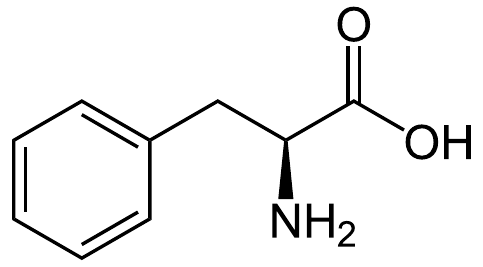}} &\ch{C9NH11O2} & 235 & 32 & 33\\
       Cysteine&\raisebox{-0.2cm}{\includegraphics[width=1.3cm]{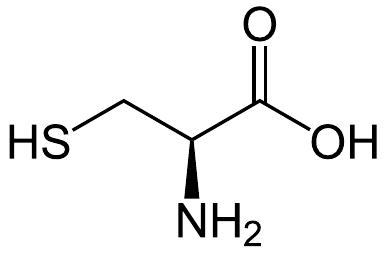}} &\ch{C3NH7SO2} & 144 & 21 & 24\\\bottomrule
     \end{tabular}
\end{table}
\begin{figure*}[t]
\centering  \includegraphics[width=0.9\textwidth]{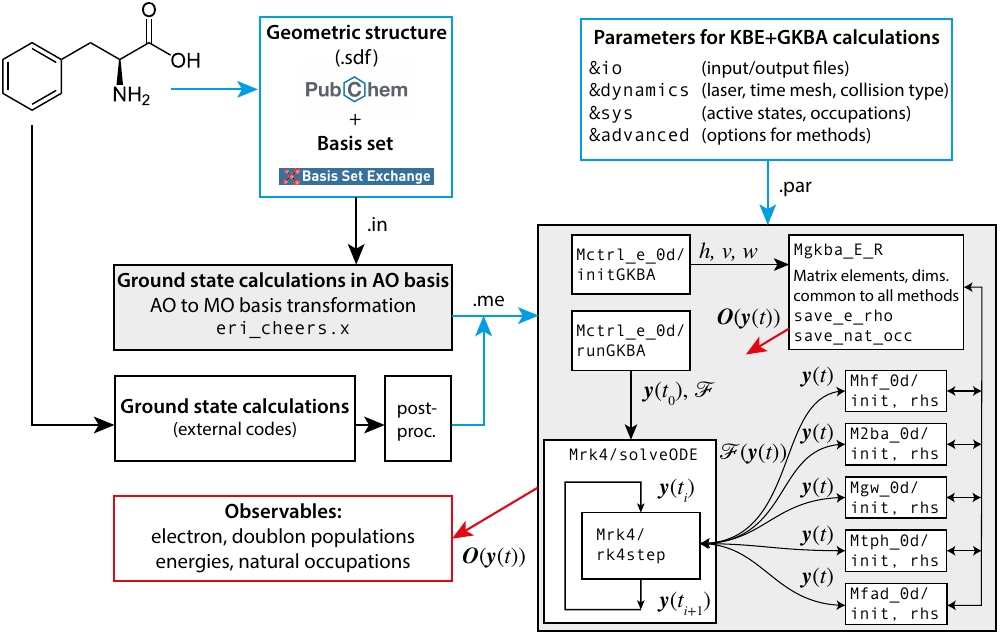}
\caption[]{Workflow diagram of the GKBA+ODE calculations of the nonequilibrium dynamics in
  \emph{molecular systems} (one can think of them as zero dimensional systems, and this is
  reflected in suffixes \texttt{\_0d} in the names of respective modules). Pieces of code
  belonging to the \cheers{} package are gray-shaded. Input parameters are framed in blue
  color, output observables are framed in red color. Most computationally intensive
  subroutines are \texttt{rhs} belonging to the modules \texttt{Mhf}, \texttt{M2ba},
  \texttt{Mgw}, \texttt{Mtph}, \texttt{Mtpp}, and \texttt{Mfad}. They all have access to
  the module \texttt{Mgkba\_E\_R}, which contains all relevant system parameters needed
  for the time-propagation. \texttt{Mrk4} is our implementation of the RK4 algorithms. It
  contains a stepper \texttt{rk4step} and a driver \texttt{solveODE}, which loops over the
  times, estimates propagation error and calls a subroutine for the computation of
  observables. Due to its simplicity and efficiency \texttt{rk4step} is not parallelized,
  whereas \texttt{rhs} are perfect candidates for parallelization. \label{fig:cheers:0d}}
\end{figure*}

\begin{figure*}[t]
\centering  \includegraphics[width=0.95\textwidth]{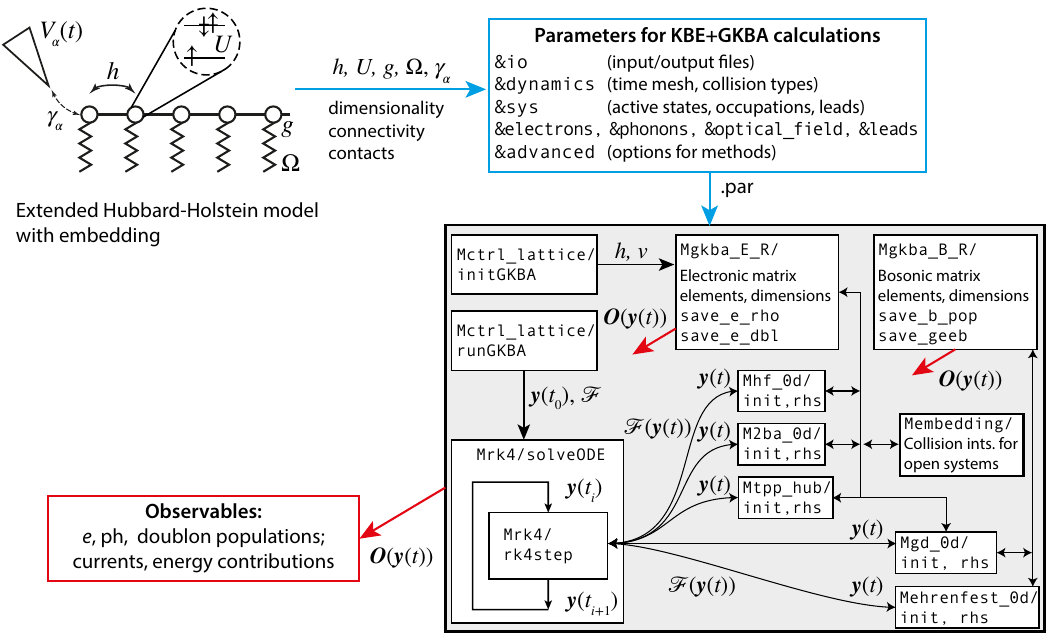}
\caption[]{Workflow diagram of the GKBA+ODE calculations of the nonequilibrium dynamics in
  \emph{model lattice systems}. Notations as in Fig.~\ref{fig:cheers:0d}. Additionally we
  have a module for storing bosonic variables \texttt{Mgkba\_B\_R} and computational
  bosonic modules \texttt{Mehrenfest\_0d} and \texttt{Mgd\_0d}. Notice that electronic and
  bosonic modules are not completely independent. For instance \texttt{Mgd\_0d} needs to
  fetch effective electronic one-body Hamiltonian stored in \texttt{Mgkba\_E\_R}.
  \label{fig:cheers:lattice}}
\end{figure*}
We focus on the scenario where several copies of the state vector $\vec{y}(t)$ can be
stored in the main memory of a computer or be distributed to each node of a computer
cluster. This is by no means obvious: typical DFT implementations completely avoid the
storage of quantities scaling as a forth power of the system size (such as Coulomb
integrals). They operate with one-body densities with quadratic scaling, whereas the
Coulomb matrix elements in atomic orbital (AO) basis are computed ``on the fly'', allowing
one to perform self-consistent calculations even for
proteins~\cite{cole_applications_2016}.

This approach is not feasible for a KBE+GKBA theory, which is often formulated in the molecular orbital (MO) basis
making the re-computation of the Hamiltonian matrix elements at each time-step infeasible. Intricacies of the basis
selection and the basis transformation are discussed by \cite{perfetto_first-principles_2019}, where it is shown how to
deal with a continuum of unbound states that participate in the photoionization processes.  This constitutes a separate
preparatory step. Of course, the fact that $N_v=\mathcal{O}\mleft(N_\text{$e$-basis}^4\mright)$ puts strong restrictions
of the feasible system sizes. Let us consider a few representative examples (Tab.~\ref{tab:aminoacids}). As can be seen
from the table, the number of atomic basis functions in the full-electron calculations is significantly larger than the
number of occupied and virtual MO states used in the GKBA+ODE dynamics. This is achieved by excluding the core and the
unoccupied states with large positive energy.  This restriction is not mathematically but physically motivated
considering typical optical excitations using an XUV laser. For instance, in ~\cite{pavlyukh_photoinduced_2021} we
performed calculations for the Glycine molecule using only 5 relevant valence states out of 15 available. This selection
was based on the insight provided by experimental measurements~\cite{weinkauf_highly_1996} and theoretical
calculations~\cite{kuleff_multielectron_2005, ayuso_ultrafast_2017}.

As a rule, in our calculations we use experimental molecular geometries from the
\textsc{PubChem}~\cite{kim_pubchem_2021} online database and Gaussian basis sets from the Basis Set
Exchange~\cite{pritchard_new_2019} online resource. We note, however, that \cheers{} is not restricted to this
scenario, it is conceptualized to work with any programs that can generate MO integrals (Fig.~\ref{fig:cheers:0d}) as
well as with model systems (Fig.~\ref{fig:cheers:lattice}). In order to get familiar with the second possibility, in
Tab.~\ref{tab:model} we present typical resources in order to study the electron localization~\cite{kloss_multiset_2019}
in the Hubbard-Holstein model. A novel feature of these calculations is that both $e$-$e$ and $e$-ph interactions are
taken into account.

\begin{table}[b]
  \caption{\label{tab:model} Largest systems and computational resources (Xeon Gold 5218 CPU @ 2.30GHz) required to
    study the dynamics of the Holstein-Hubbard model with $N$ sites and physical propagation time $t_\text{f}$. We
    compare timings and sizes of the state vector for a set of electronic (HF\,---\,Hartree-Fock, 2B\,---\,second Born,
    $T^{pp}$\,---\,$T$-matrix in the particle-particle channel) and bosonic (Ehrenfest\,---\,mean-field,
    $GD$\,---\,Fan-Migdal approximation) methods. } \arrayrulecolor{NavyBlue}
    \begin{tabular}{cccrccc}\toprule
      System & \multicolumn{2}{c}{Correlations} & State & Time & \multicolumn{2}{c}{CPU hours}\\\cline{2-3}\cline{6-7}
      $N$& $e$-$e$  & $e$-ph& vector & $t_\text{f}$ &\mbox{$e$-$e$}  &\mbox{$e$-ph}\\\midrule \renewcommand{\arraystretch}{1.4}
      
      151         & HF & Ehrenfest &   23\,254   &  40 &0.02   & 0.07\\
      151         & HF & $GD$      & 7\,046\,264   &  40 &3.8    &  2.0\\
      151         & 2B & $GD$      & 526\,954\,666 &  40 &181.2  &  2.0\\
      151         & $T^{pp}$ & Ehrenfest & 519\,931\,656 &40 & 482.8 & 0.04\\
      151         & $T^{pp}$ & $GD$ & 526\,954\,666 &  40 &488.0 &   2.6\\\bottomrule
       \end{tabular}
\end{table}

It can be concluded on the basis of these two examples that GKBA+ODE calculations are routinely possible for amino acid
molecules and correlated 1D clusters. From the structure of the code depicted in
Figs.~\ref{fig:cheers:0d},\,\ref{fig:cheers:lattice}, the separation of the ODE solver from the data generation and the
RHS evaluation ($\mathcal{F}\mleft\{t, \vec{y}(t)\mright\}$) represents an optimal architecture amenable to different
parallelizations. It is anticipated that our code can be applied to a variety of physical scenarios, including ultrafast
charge migration in XUV photoexcited halogenated amino acids~\cite{bergamaschi_halogen_2018}, 2D and 3D model
systems. In the next sections we consider some representative blocks of code according to the workflow diagrams. More
detailed description of the full structure is deferred to the Supporting Information.

\subsection{Ground state properties and the initialization of the ODE solver\label{sec:init}}
Let us start with molecular systems. As can be seen from the workflow in Fig.~\ref{fig:cheers:0d}, a ground state
calculation is required in order to generate matrix elements for subsequent propagation. They are stored in a binary
format in the file with an extension \texttt{.me} containing the following records:
\begin{description}
\item[\texttt{me1p}] One-particle matrix elements. They typically comprise the kinetic energy operator, electron-nuclear
  repulsion and the mean-field interaction of valence electrons with the core electrons. Depending on the method, core
  electrons may or may not be included in the ground state calculation. The active space for dynamical calculations is
  determined by physical considerations.
\item[\texttt{medp}] Dipole transition matrix elements along three Cartesian axes.
\item[\texttt{me2p}] Two-particle matrix elements representing the Coulomb repulsion
  between the electrons.
\end{description}  
Matrix elements are assumed to be real and they are stored in a \emph{packed form} in accordance with their symmetries.
One of the first tasks is to read the matrix elements (\texttt{me1p}, \texttt{medp}, \texttt{me2p}) and the system
dimensions (the number of electronic states \texttt{NP} and highest occupied molecular orbital \texttt{NF}) and store
them in the \texttt{Msys\_0d} module. We typically assume that HF Hamiltonian is diagonal in MO basis. This is dictated
by practical rather than physical considerations: for instance one may study correlated molecular dynamics starting from
a weakly correlated ground state in which the initial density matrix is diagonal and stationary.

However, \cheers{} is not limited to molecular systems. Correspondingly, there are
several \texttt{Msys\_} modules. Information on the \emph{lattice systems} is stored in
the \texttt{Msys\_lattice} module, whereas for restricted Hartree-Fock calculations in
\emph{atomic basis}, the \texttt{eri\_cheers.x} program (part of \cheers) makes use of the
\texttt{Msys\_molecule} module to store the information about the AO-basis.

The second input file, which is mandatory for all GKBA+ODE calculations, is a parameter file \texttt{.par}. It consists of
several \textsc{namelists} reflecting different aspects of a simulation:
\begin{description}
\item[\texttt{io}] provides names of the matrix elements file and output files for observables.
\item[\texttt{dynamics}] specifies the propagation time-interval, time-step, and the level of treatment of electronic
  and bosonic correlations. Currently, only a fixed time-step ODE solver is used. Although it may not be the fastest
  option, this choice ensures that calculations remain reproducible and directly comparable to other codes, thereby
  simplifying the debugging process.
\item[\texttt{sys}] gives MO basis size, index of highest occupied molecular orbital, and dimensionality, number of
  orbitals per site, number of leads for lattice systems. The number of read MO states (\verb|np|) may in general be
  different (smaller or equal) from the number of MO states in the \verb|.me| file. Also, \verb|np| may in general be
  different (greater or equal) from the size of electronic basis for the GKBA+ODE calculations, $N_\text{$e$-basis}$
  (\verb|ns|). This gives one additional flexibility to experiment with different sets of active molecular orbitals and
  to exclude irrelevant valence states from the consideration (known in advance to be always occupied or always empty).
\item[\texttt{electrons}] provides details of electron hopping (nearest and next-nearest) and Coulomb interaction,
  periodicity, spinless option, and extension for lattice systems. The resulting electronic quantities are then stored
  in \texttt{Mgkba\_E\_R}. This is one of the most important modules as it contains all the data for electronic GKBA+ODE
  calculations. As is depicted in Fig.~\ref{fig:cheers:0d}, the computational routines in \texttt{Mhf}, \texttt{M2ba},
  \texttt{Mgw}, \texttt{Mtph}, \texttt{Mtpp}, and \texttt{Mfad} make use of the matrix elements stored in
  \texttt{Mgkba\_E\_R} and also use the module for intermediate quantities.
\item[\texttt{phonons}] gives phononic frequencies and parameters of $e$-ph interaction. Based on them, corresponding
  matrix elements are constructed and stored in the \texttt{Mgkba\_B\_R} module. The computational routines in the
  bosonic modules \texttt{Mehrenfest\_0d} and \texttt{Mgd\_0d} make use of the matrix elements and also use the module
  for the storage of intermediate quantities.
\item[\texttt{optical\_field}] contains parameters of the laser pulses, such as duration, frequency and shape.
\item[\texttt{leads}] contains tunneling rate matrices for the wide band limit approximation (WBLA), or details of the
  leads and tunneling Hamiltonians;
\item[\texttt{advanced}] specifies additional options for the methods used, such as, inclusion of exchange, freezing of
  mean field Hamiltonians, presence of acoustic phonons, treatment of Coulomb integrals with two indices or as complex
  quantities. Each of the methods may have different options, all the possibilities are described in \texttt{Moptions}.
\end{description}  
\subsection{Organization of the state vector}
\begin{figure}[]
\centering  \includegraphics[width=5cm]{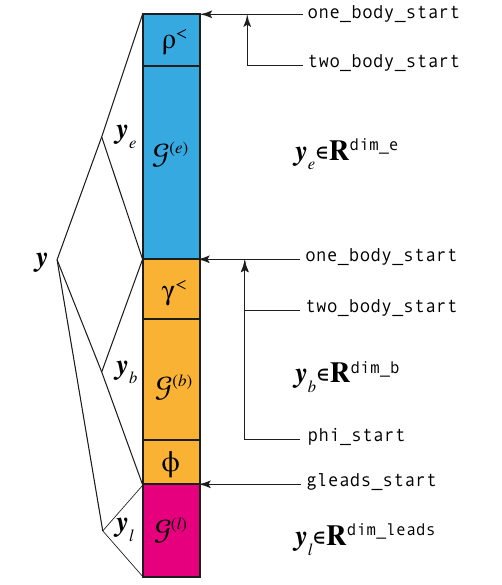}
\caption[]{Organization of the state vector. Electronic, bosonic and leads part with dimensions
  \texttt{dim\_e}, \texttt{dim\_b}, and \texttt{dim\_l}, respectively, are indicated with different colors. In the electronic part
  the electron density matrix $\rho^<$ (Eq.~\ref{eq:def:rho}) and two-electron correlator $\cG^{e}$ (Eq.~\ref{def:G:e})
  can be accessed at off-sets \texttt{one\_body\_start} and \texttt{two\_body\_start}, respectively. In the bosonic
  part, the bosonic density matrix $\gamma^<$ (Eq.~\ref{eq:def:gamma}), electron-boson correlator $\cG^{b}$
  (Eq.~\ref{def:G:b}), and expectation values of bosonic field operators $\phi$ (Eq.~\ref{def:phi}) can be accessed at 
 off-sets \texttt{one\_body\_start}, \texttt{two\_body\_start} and \texttt{phi\_start}, respectively. Finally, the
 embedding correlator $\cG^\text{em}$ (Eq.~\ref{eq:emcorr}) can be found at the off-set \texttt{gleads\_start}.\label{fig:state}}
\end{figure}

The code comprises several executables (such as shown in
Figs.~\ref{fig:cheers:0d},\,\ref{fig:cheers:lattice}). Therefore, the state vector $\vec{y}$ is physically located in
different modules \texttt{Mctrl\_e\_0d} (molecular systems), \texttt{Mctrl\_lattice} (lattice systems) and in
\texttt{Mctrl\_transport} (lattice systems with embedding).  It always contains the electronic component, and can
optionally contain the phononic (or bosonic) and embedding (or leads) parts, i.e.,
\begin{align}
  \vec{y}(t)&=\{\vec{y}_e(t),\vec{y}_b(t),\vec{y}_l(t)\},
  \label{eq:y}
\end{align}
as depicted in Fig.~\ref{fig:state}. In order to properly define each component, let us consider the most general system
Hamiltonian.
\paragraph{System Hamiltonian}
\begin{equation}\label{eq:totalHamiltonian}
 \Hh(t) = \Hh_\el(t) + \Hh_\bos(t) +\Hh_\elbos(t)+\Hh_{\text{leads}} + \Hh_{\text{tunnel}}.
\end{equation}
It contains the electronic part
\begin{align}\label{eq:H:e}
 \Hh_\el(t)&=\sum_{ij} h_{ij}(t)\hd_{i}^\dagger \hd_{j}
+\frac12\sum_{ijmn}v_{ijmn}(t)\hd_{i}^\dagger \hd_{j}^\dagger \hd_{m} \hd_{n},
\end{align}
comprising a one-body term ($h^\dagger=h$) accounting for the kinetic energy as well as the interaction with nuclei and
possible external fields. The time-dependence of $h_{ij}(t)=h^{0}_{ij}+\bra i|V_{\rm
  gate}(t)+\hat{\mathbf{d}}\cdot{\mathbf{E}_{\rm pulse}}(t)|j\ket$ is due to a time-dependent gate voltage
$V_{\mathrm{gate}}(t)$ and to a possible laser pulse $\mathbf{E}_{\rm pulse}(t)$ coupled to the electronic dipole
operator $\hat{\mathbf{d}}$. Furthermore, there is a two-body term accounting for the Coulomb interaction between the
electrons. The time-dependence of the Coulomb matrix elements $v_{ijmn}(t)=v_{ijmn}s(t)$ could be due to the adiabatic
switching protocol $s(t)$ adopted to generate a correlated initial state. Latin indices denote one-electron states
comprising orbital and spin degrees of freedom, i.~e., $i=(\bm{i},\sigma)$. As can be seen from
Figs.~\ref{fig:cheers:0d},\,\ref{fig:cheers:lattice}, $h$ and $v$ matrix elements are stored in the \texttt{Mgkba\_E\_R}
module.

We write the  bosonic Hamiltonian as
\begin{align}\label{eq:H:b}
 \Hh_\bos(t)= \sum_{\mu\nu} \Omega_{\mu\nu}(t)\hphi_\mu\hphi_\nu,
\end{align}
where $\Omega^\dagger =\Omega$ may depend on time, such as in the parametric phonon
drivings~\cite{murkakami_nonequilibrium_2017} setup. The annihilation and creation operators for a bosonic mode $\bgm$,
i.e., $\hat{a}_{\fbgm}$ and $\hat{a}^{\dag}_{\fbgm}$, are arranged into a vector $(\hat{x}_{\fbgm},\hat{p}_{\fbgm})$
where $\hat{x}_{\fbgm}=(\hat{a}^{\dag}_{\fbgm}+\hat{a}_{\fbgm})/\sqrt{2}$ are the position operators and
$\hat{p}_{\fbgm}=i(\hat{a}^{\dag}_{\fbgm}-\hat{a}_{\fbgm})/\sqrt{2}$ are the momentum operators.  The greek index
$\mu=(\bgm,\xi)$ is then used to specify the bosonic mode and the component of the vector: $\hphi_{\mu}=\hat{x}_{\fbgm}$
for $\xi=1$ and $\hphi_{\mu}=\hat{p}_{\fbgm}$ for $\xi=2$. The electronic and bosonic subsystems interact through
\begin{align}\label{eq:H:e:b}
 \Hh_\elbos(t)&= \sum_{\mu, ij} g_{\mu, ij}(t)\hd_{i}^\dagger 
 \hd_{j}\hphi_\mu;
\end{align}
therefore electrons can be coupled to both the mode coordinates and momenta.  We allow $g$ to depend on time for
possible adiabatic switchings, i.\,e., $g_{\mu, ij}(t)=g_{\mu, ij}s(t)$. $\Omega$ and $g$ matrix elements are stored in the
\texttt{Mgkba\_B\_R} module.

Similarly to the bosonic part, the lead part of the total Hamiltonian is optional: not every calculation is performed
with transport setup in mind, moreover the leads and the tunneling Hamiltonian parts do not explicitly enter the
formalism in the wide-band limit approximation. Nonetheless, we provide some possible explicit forms of these parts in order to
give a microscopic definition of the tunneling-rate matrices. Thus, the leads Hamiltonian can be parametrized as
\begin{align}
  \Hh_{\text{leads}} &= \sum_{k, \alpha,\sigma}\epsilon_{k \alpha}\hc_{k \alpha,\sigma}^\dagger \hc_{k \alpha,\sigma}.
\end{align}
At the moment we assume one-dimensional momenta $k$. Here $\alpha$ denotes the lead number and $\sigma$ is the spin
projection. For a one-dimensional tight-binding chain (with on-site energy $a_\alpha$ and hopping $b_\alpha$) the
electron energy dispersion $\epsilon_{k\alpha}$ is parametrized as
\begin{align}
  \epsilon_{k \alpha,\sigma}&=a_\alpha + 2|b_\alpha| \cos\mleft(\frac{k\pi}{N_k+1}\mright),&
  1&\le k \le N_k.\label{eq:k:disp}
\end{align}
As can be seen, $a$ and $b$ are lead-dependent, whereas we assume the same number of momenta $N_k$ per each lead. Here,
leads are not spin-polarized, but it would be straightforward to consider, e.\,g., ferromagnetic leads via modified lead
energies. Upon the application of a time-dependent bias voltage $V_\alpha(t)$, the lead energies are modified as
\begin{align}
  \tilde{\epsilon}_{k \alpha,\sigma}(t)&= \epsilon_{k \alpha,\sigma} + V_\alpha(t).\label{eq:k:disp:bias}
\end{align}

The tunneling Hamiltonian can be parametrized as:
\begin{align}
  \Hh_{\text{tunnel}}(t) &= \sum_{k, \alpha}\sum_{\bm{i},\sigma}T_{\bm{i},k\alpha}(t)
  \left[\hd_{\bm{i},\sigma}^\dagger \hc_{k \alpha,\sigma}+h.c.\right].\label{eq:H:tunnel}
\end{align}
The coupling strengths are in general time- and lead-dependent, but it is assumed that for
each lead $\alpha$ there is a \emph{single contact site}, i.e.,
\begin{align}
  T_{\bm{i},k\alpha}(t)&=t_\alpha\delta_{\bm{i} C_\alpha}\sqrt{\frac{2}{N_k+1}}\sin\mleft(\frac{k\pi}{N_k+1}\mright) s_\alpha(t).
  \label{eq:T:def}
\end{align}
$C_\alpha$ represents the quantum number (such as site number and band index) of the electronic state that couples to
the lead $\alpha$. $s_\alpha(t)$ is the switch-on function between the system and the lead $\alpha$, see
Fig.~\ref{fig:cheers:lattice}. Next we define the tunneling-rate matrix
\begin{align}\label{eq:tunneling-rate-matrix}
\Gamma_{\alpha,\bm{ij}}(\omega; t) & = 2\pi\sum_{k=1}^{N_k} T_{\bm{i},k\alpha}(t)\delta(\omega-\epsilon_{k\alpha})T_{\bm{j},k\alpha}^*(t),
\end{align}
which is in general frequency-dependent, but in WBLA it reduces to
\begin{align}
\Gamma_{\alpha,\bm{ij}}(\omega; t) & = \gamma_{\alpha}\delta_{\bm{i} C_\alpha}\delta_{\bm{j} C_\alpha} s_\alpha^2(t).
\end{align}

\paragraph{Parts of the state vector}
The electronic component $\vec{y}_e$ contains i) the electron density matrix
\begin{align}
  \rho_{ij}^<&=-iG^{<}_{ij}(t,t),\label{eq:def:rho}\\
  G^{<}_{ij}(t,t')&=i\langle \hd^\dagger_j(t')\hd_i(t)\rangle,
\end{align}
proportional to the equal-time lesser Green's function $G^{<}$; and ii) a two-electron correlator
\begin{align}
\cG^{e}_{imjn}(t)&=-\langle 
\hat{d}^{\dag}_{n}(t)\hat{d}^{\dag}_{j}(t)\hat{d}_{i}(t)\hat{d}_{m}(t)\rangle_{c}, \label{def:G:e}
\end{align}
The subscript ``$c$'' in the averages signifies that only the correlated part must be retained.

The bosonic component $\vec{y}_b$ contains i) the bosonic density matrix $\gamma^<_{\mu\nu}$
\begin{align}
  \gamma_{\mu\nu}^{<}(t)&= i D^{<}_{\mu\nu}(t, t),\label{eq:def:gamma}\\
   D^<_{\mu \nu}(t,t')&=D^>_{\nu\mu}(t',t) =
   -i \langle \Delta \hphi_{\nu}(t') \Delta \hphi_{\mu} (t) \rangle,\label{eq:def:D}\\
   \Delta \hphi_{\nu}(t)&\equiv\hphi_{\nu}(t) -\langle \hphi_{\nu}(t)\rangle,\label{eq:def:dphi}
\end{align}
proportional to the equal-time lesser bosonic Green's function $D^{<}_{\mu\nu}$; ii) a higher-order electron-boson
correlator
\begin{align}
 \cG^{b}_{\mu,ij}(t)&=\langle\hat{d}^{\dag}_{j}(t)\hat{d}_{i}(t)\hat{\phi}_{\mu}(t)\rangle_{c},\label{def:G:b}
\end{align}
and iii) the expectation values of the bosonic field operators:
\begin{align}
\phi_\nu(t)&\equiv\langle \hphi_{\nu}(t)\rangle.\label{def:phi}
\end{align}
Matrices $\gamma^<$, $D^{<}$ have dimensions $N_\text{$b$-basis}\times N_\text{$b$-basis}$.

To deal with open systems, the leads part $\vec{y}_l$ contains the embedding correlator $\cG^\text{em}$, which can be
completely specified in terms of the leads and tunneling Hamiltonians, together with the quasi-particle propagators of
the coupled system. This is due to the leads being non-interacting and in contrast to other higher-order correlators
such as Eq.~\eqref{def:G:e}. Here, we consider the wide-band limit approximation by expressing the embedding correlator
within the so-called pole expansion scheme~\cite{tuovinen_time-linear_2022},
\begin{align}
  \mathcal{G}_{l\alpha}^{\text{em}}(t) = \int d\bar{t} s_\alpha(\bar{t})
  \ex^{-i\phi_\alpha(t,\bar{t})}\ex^{-i\mu(t-\bar{t})}\ex^{-\zeta_l(t-\bar{t})/\beta} G^{\text{A}}(\bar{t},t) .\label{eq:emcorr}
\end{align}
While we will next specify all the components of Eq.~\eqref{eq:emcorr}, we refer the Reader to the Supporting
Information and our recent letter~\cite{tuovinen_time-linear_2022} for additional details. The embedding correlator,
besides depending on the lead index $1\le\alpha\le N_\text{leads}$ contains the dependence on the pole index $1\le l\le
N_p$ in the expansion of the Fermi distribution function
\begin{align}
  f(x)=\frac{1}{\ex^{\b x}+1} &= \frac{1}{2} - \lim_{N_p\to \infty}
  \sum_{l=1}^{N_p} \eta_l \left(\frac{1}{\b x+i\zeta_l}+\frac{1}{\b x -i\zeta_l}\right) ,
\end{align}
where $\eta$ and $\pm i\zeta$ are the residues and poles ($\zeta>0$), respectively. We typically use the expansion
coefficients generated by solving an eigenvalue problem of a specific, tridiagonal
matrix~\cite{hu_communication:_2010}. Additionally $\phi_\alpha(t,t')$ is the bias-voltage phase factor
\begin{align}
  \phi_\alpha(t,t') &\equiv \int_{t'}^t d\bar{t}V_\alpha(\bar{t}),
\end{align}
$\mu$ is the chemical potential, and
\begin{align}
  G^{\text{R}/\text{A}}(t,t') = \mp i \theta[\pm(t-t')]\mathrm{T} \ex^{-i\int_{t'}^t d\bar{t} h_\text{eff}^e(\bar{t})}
\end{align}
are the quasiparticle retarded/advanced propagators of the coupled system. Notice, that at variance with closed systems,
the effective Hamiltonian is in general non-self-adjoint
\begin{align}
h_\text{eff}^e(t)&= h^e(t) - \frac{i}{2}\sum_\alpha \Gamma_\alpha s_\alpha^2(t)= h^e(t)- \frac{i}{2}\Gamma(t).
\end{align}
Matrices $h^e$, $\Gamma$, $G^{\text{R}/\text{A}}$, $\mathcal{G}_{l\alpha}^{\text{em}}$ have dimensions
$N_\text{$e$-basis}\times N_\text{$e$-basis}$.  We conclude this section by defining the mean-field electronic
Hamiltonian
\begin{align}
  h^e_{ij}(t)&=h_{ij}(t)+V^{\rm HF}_{ij}(t),\\
  V^{\rm HF}_{ij}(t)&=\sum_{mn}w_{imnj} \rho^<_{nm}(t),
\end{align}
with $w_{imnj}=v_{imnj}-v_{imjn}$ being the antisymmetrized interaction. The storage space for
$\vec{y}$\,---\,$\mathcal{O}(N_v)$\,---\,is dominated by the $\cG^{e}$
correlator\,---\,$\mathcal{O}(N_\text{$e$-basis}^4)$, the electron-boson correlator
$\cG^{b}$\,---\,$\mathcal{O}(N_\text{$e$-basis}^2N_\text{$b$-basis})$, and the embedding
correlator\,---\,$\mathcal{O}(N_\text{$e$-basis}^2N_\text{leads}N_p)$, where $N_\text{$b$-basis}$ is double of the
number of bosonic modes, and the number of poles in the expansion of the Fermi distribution function is typically below
50.
\subsection{Computational subroutines}
In the previous section we introduced the electronic, bosonic and leads parts of the state vector. Equations of motion
covering the combinations of electronic-bosonic and electronic-leads degrees of freedom were derived in
Ref.~\cite{pavlyukh_time-linear_2022-1} and Ref.~\cite{tuovinen_time-linear_2022}, respectively. These equations
demonstrate the conserving properties~\cite{karlsson_fast_2021} and symmetries of underlying theories. It follows, for
instance, that density matrices $\rho^<$ (Eq.~\ref{eq:def:rho}) and $\gamma^<$ (Eq.~\ref{eq:def:gamma}) are self-adjoint
and that higher-order correlators can be likewise written in matrix form, by combing electronic indices in pairs. From
the electronic correlator tensor $\cG^{e}_{imjn}$ (Eq.~\ref{def:G:e}) a matrix $\mcG^{e}(t)$ can be constructed in three
different ways depending on the approximation used: $GW$, $T^{ph}$, or $T^{pp}$. These approximations contain an
infinite sequence of diagrams corresponding to the solution of the Bethe-Salpeter equation in three different
channels. $GW$ approximation corresponds to the treatment of $\overline{ph}$ channel, $T^{pp}$ approximation\,---\,$pp$
channel, and $T^{ph}$\,---\,$ph$ channel, where we refer to the naming convention used in the parquet
theory~\cite{pavarini_dynamical_2014}. Additionally, it is possible to take even more complicated diagrams into account
by including exchange contributions~\cite{pavlyukh_time-linear_2022-1}. In every such form the matrix $\mcG^{e}(t)$ can
be shown to be self-adjoint~\cite{pavlyukh_photoinduced_2021}. This matrix form of the GKBA+ODE method leads to
important simplifications of the resulting equations, for instance only the upper triangular parts must be computed. For
the explicit form of the equations we refer to corresponding
papers~\cite{pavlyukh_time-linear_2022-1,tuovinen_time-linear_2022} and focus here more on their general functional
form.

The evaluation of $\mathcal{F}\mleft\{t, \vec{y}(t)\mright\}$ in Eq.~\eqref{eq:ode}, which constitutes the main
computational task, can be partitioned as
\begin{align}
  \mathcal{F}\mleft\{t, \vec{y}(t)\mright\}&=
  \left\{\mathcal{F}_e\mleft\{t, \vec{y}_e(t),\vec{y}_b(t),\vec{y}_l(t)\mright\},
  \mathcal{F}_b\mleft\{t, \vec{y}_e(t),\vec{y}_b(t)\mright\},
  \mathcal{F}_l\mleft\{t, \vec{y}_e(t),\vec{y}_l(t)\mright\}
  \right\}.
\end{align}  
As can be seen from this form, the evaluation of one part requires the knowledge of other parts. This structure is
present even at the mean-field level, where the equation of motion for the bosonic and leads correlators depends on the
electron density matrix. In correlated cases, there are feedbacks of these degrees of freedom on electrons, and the
electron density matrix $\rho^<$ enters the evaluation of the higher-order correlators. These complicated relations are
shown as arrows in Figs.~\ref{fig:cheers:0d},\,\ref{fig:cheers:lattice}. The subroutines implementing different
electronic, bosonic and embedding approximations are grouped in modules such as:
\begin{description}
\item[\texttt{Mhf\_0d}]  mean-field for electrons,
\item[\texttt{M2ba\_0d}] second Born approximation for electrons,
\item[\texttt{Mgw\_0d} and \texttt{Mgw\_hub}] $GW$ approximation for molecular and lattice systems,
\item[\texttt{Mtpp\_0d} and \texttt{Mtpp\_hub}] $T^{pp}$ approximation for molecular and lattice systems,
\item[\texttt{Mtph\_0d} and \texttt{Mtph\_hub}] $T^{ph}$ approximation for molecular and lattice systems,
\item[\texttt{Mfad\_0d}] three-particle Faddeev method for molecular systems,
\item[\texttt{Mehrenfest\_0d} and \texttt{Mgd\_0d}] Ehrenfest and Fan-Migdal (or $GD$) approximation for phonons,
\item[\texttt{Membedding}] the spectral decomposition and the pole expansion methods for leads.
\end{description}

In addition to the state vector, the \texttt{rhs} subroutines (computing right hand sides of the equations of motion) in
these modules rely on the common data (matrix elements) stored in \texttt{Mgkba\_E\_R} and \texttt{Mgkba\_B\_R}
modules. The code is organized in such a way that subroutines implementing different approximations are completely
interchangeable. For instance, every evaluation starts with a transformation between the state vector (real) $\vec y$
and the matrix form (complex) of $\rho^<$, $\mcG^{e}(t)$ for electrons, $\gamma^<$, $\mcG^{b}(t)$, $\phi$ for bosons,
and $\mathcal{G}^{\text{em}}$ for leads. This computational overhead is needed for three reasons: i) simplicity and
universality of the differential equation solver \texttt{solveODE}; ii) speed-up due to automatic loops unrolling in the
Runge-Kutta algorithm, \texttt{rk4step}; iii) the computational complexity of the transformation operations is well
below time-critical operations, which scale as fifth or higher power of the system size. There is also small memory
saving due to the use of symmetry, which may play an important role in distributed computing. Therefore, it is possible
to achieve a great variety of methods without code duplication. The details of this design are presented in Supporting
Information. Let us now demonstrate the interplay of electronic correlations, electron-phonon interaction and tunneling
to the leads in a model calculation.
\section{Illustrative example}
We consider here a 1D Hubbard model at low filling interacting at every site with a phonon mode. This is also known as
the Holstein-Hubbard model, which is often studied in the context of polaron formation and electron
localization~\cite{kloss_multiset_2019} and field-driven heating~\cite{weber_electronic_2023}. We add here another
aspect to these studies, namely the presence of leads connected to the two terminal sites ($N_\text{leads}=2$). This
designates our model as a prototype for a photovoltaic device encompassing the phenomena of creation of nonequilibrium
carriers, formation of the polaronic quasiparticles, and charge transport through the active material to leads. We focus
here only on modelling of the conduction band. A realistic model of the whole photovoltaic device should also comprise
the valence band, from where the electrons are excited, and the optical pulse. This calculation is fully within the
capability of the code. However, here we deal with a simpler one-band scenario and perform calculations for shorter
chains as compared to previous studies~\cite{pavlyukh_time-linear_2022}\,---\,only 35 sites ($N_v=1792350$)\,---\,but
they capture all relevant physical phenomena. As a benefit, all the calculations can be performed on a typical laptop
within five minutes time. The Hamiltonian is given by
\begin{align}
  \hat{H}_\text{h-h}&=-t_\text{nn}\sum_{\sigma=\uparrow,\downarrow}\sum_{<\bm{i},\bm{j}>}\hd_{\bm{i}\sigma}^\dagger \hd_{\bm{j}\sigma}
  +U\sum_{\bm{i}}\hat{n}_{\bm{i}\uparrow}\hat{n}_{\bm{i}\downarrow}
  +\sum_{\bm{i}}\left\{\Omega \ha^\dagger_{\bm{i}}\ha_{\bm{i}}+g(\ha^\dagger_{\bm{i}}+\ha_{\bm{i}})\hat{n}_{\bm{i}}\right\},
  \label{eq:h:h-h}
\end{align}
where $\hat{n}_{\bm{i}}=\sum_\sigma \hat{n}_{\bm{i}\sigma}$, $\hat{n}_{\bm{i}\sigma}=\hd_{\bm{i}\sigma}^\dagger
\hd_{\bm{i}\sigma}$, and we set $t_\text{nn}=\Omega=1$, and $U=2$, $g=0.5$. The dynamics is triggered by the creation of
an electron pair at the central lattice site. Initially the lattice is empty. In the absence of leads the electronic
wave packet spreads on the lattice, whereby the electron group velocity is at most $2t_\text{nn}$. Two electrons
occupying the same lattice site form a doublon, a long-living excitation, the decay of which is suppressed by energy
conservation.  Its group velocity is in general different from the electronic one and in the strongly interacting limit
can be estimated as $4t_\text{nn}^2/U$. Propagation of electrons and doublons at low filling can be accurately described
by the $T^{pp}$ approximation~\cite{pavlyukh_time-linear_2022}. Both velocities are renormalized by the $e$-ph
interaction.

\begin{figure}[]
  \centering
  \includegraphics[width=7.5cm]{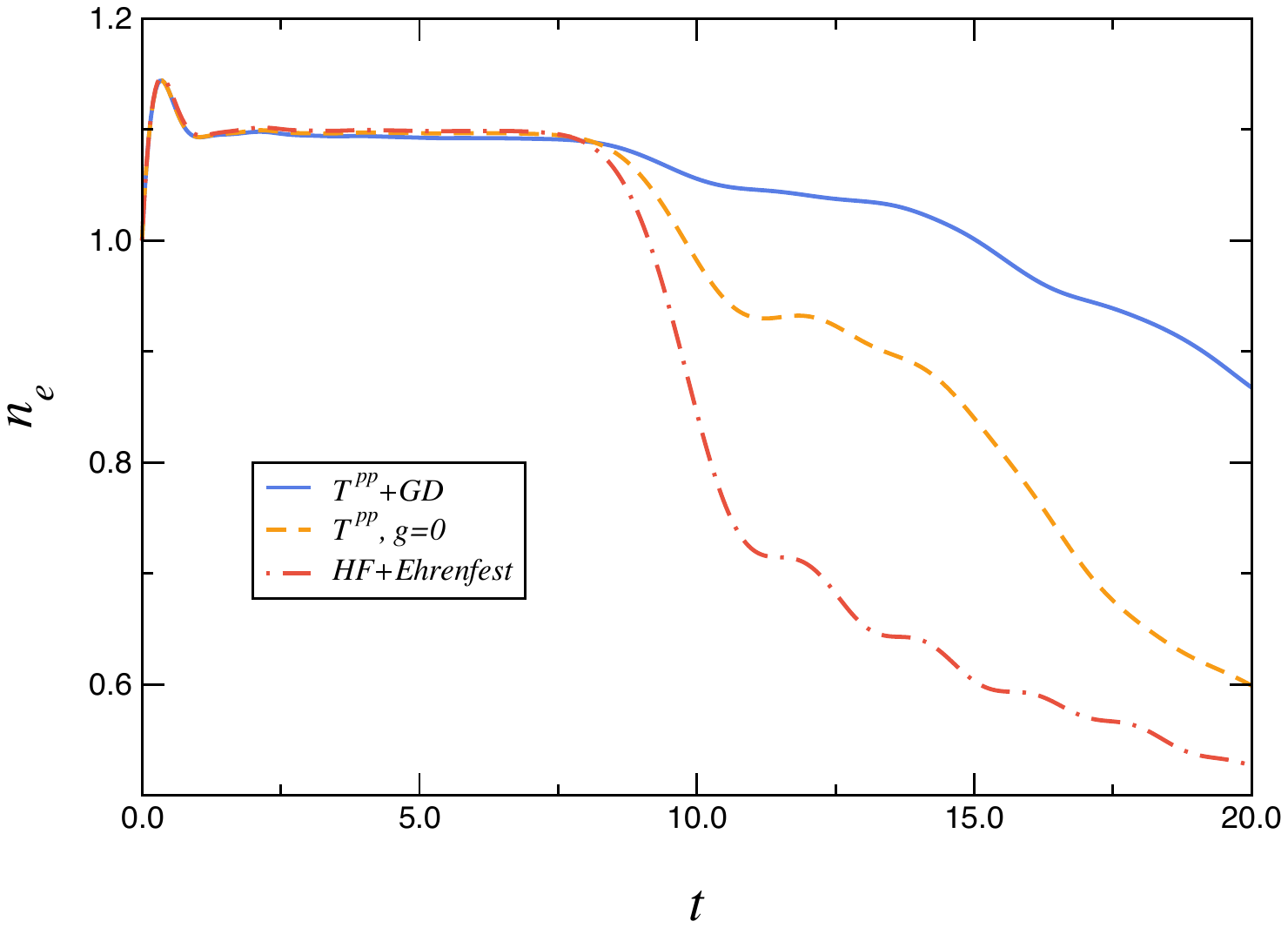}
  \includegraphics[width=7.5cm]{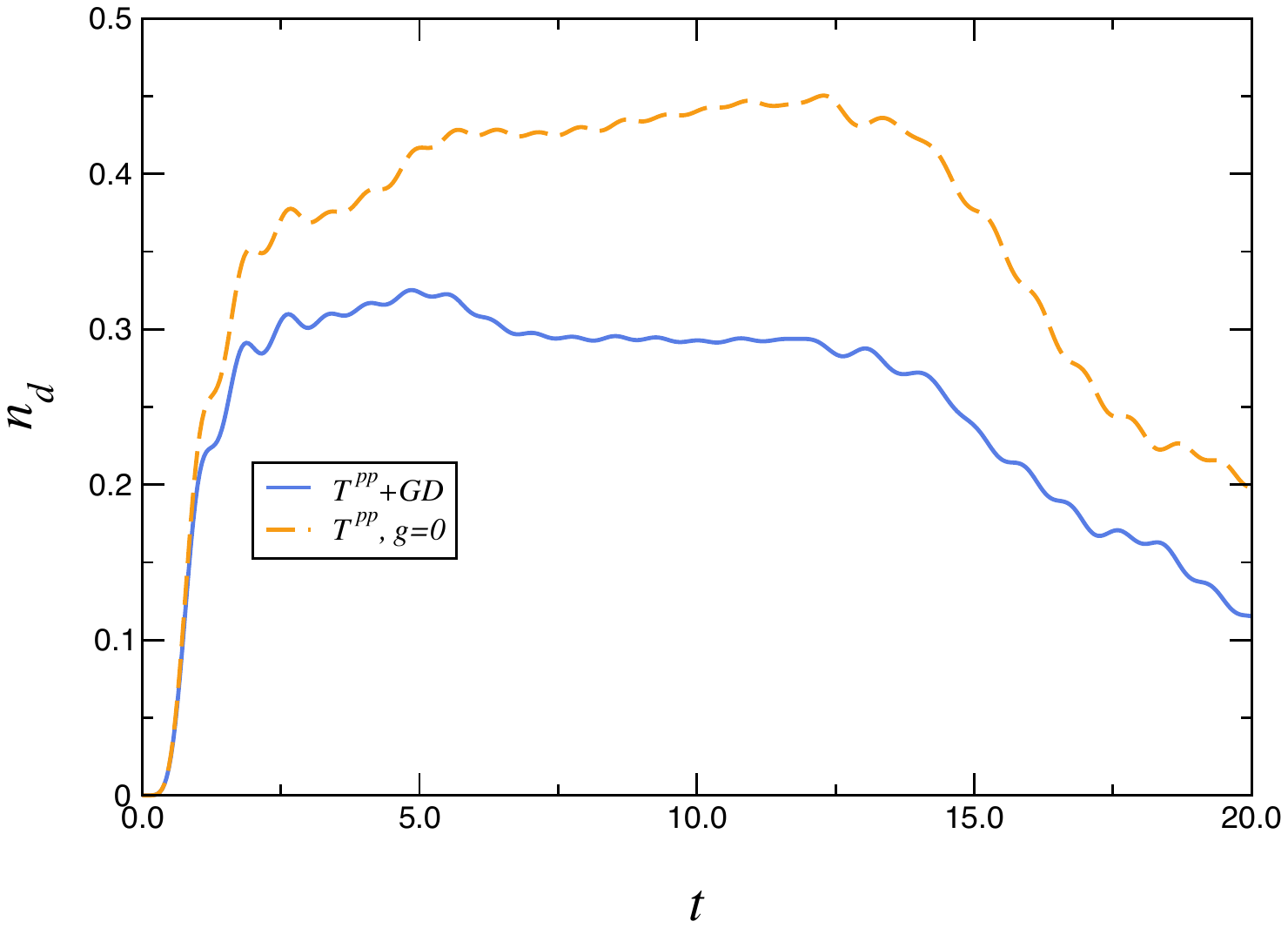}
 
\caption[]{Total number of electrons and doublons in a 35-sites system connected to leads with $V_\alpha=-5$,
  $\gamma_\alpha=1$, and $\beta=100$. Different line-styles denote different levels of treatment of $e$-$e$ and $e$-ph
  interactions. Electrons are characterized by the nearest-neighbor hopping $t_\text{nn}=1$ and on-site Hubbard
  interaction $U=2$. The phonon frequency is set to $\Omega=1$, and the electron-phonon coupling constant for
  calculations within the Fan-Migdal approach (blue solid lines) or the Ehrenfest approach (red dash-dotted line) is set
  to $g=0.5$. The orange dashed line denotes results without the electron-phonon interaction ($g=0$). Electrons are
  treated either at the Hartree-Fock (HF) or the $T$-matrix in the particle-particle channel level of
  approximation.\label{fig:1dchain_leads}}
\end{figure}

If such a system is subject to open boundary conditions, the electronic wave-packet will hit the chain boundaries and
reflect after propagating for a certain time. If the system is connected to leads, the charge will tunnel into or from
the central region, and the number of electrons in the system will not be preserved. In the present scenario, we study
how $e$-$e$ and $e$-ph interactions influence the charge flow to leads in the wide band limit approximation. The
chemical potential of the leads and central system is assumed to be equal, however, we set the lead voltage to a large
negative value ($V_\alpha=-5$). This enables electrons to leave the chain. The dynamics of the system can be understood
based on the analysis of total populations in Fig.~\ref{fig:1dchain_leads}.

The considered system is not spin polarized. Therefore, only the spatial components of $\rho^<$ and $\cG^e$ are
propagated. They are defined as
\begin{align}
  \rho_{\vec{i}\vec{j}}^<&\equiv \rho_{\vec{i}\sigma,\vec{j}\sigma}^<,\\
  \cG^e_{\vec{i}\vec{m}\vec{j}\vec{n}}&\equiv \cG^e_{\vec{i}\sigma,\vec{m}\sigma,\vec{j}\sigma,\vec{n}\sigma},
\end{align}
where $\sigma=\uparrow,\downarrow$.
The electronic and doublonic populations are defined then as respective diagonal elements 
\begin{align}
  n_{\vec{i}}&\equiv  \langle \hat{n}_{\vec{i},\sigma}\rangle =\rho_{\vec{i}\vec{i}}^<,\\
  d_{\vec{i}}&\equiv \langle \hat{n}_{\vec{i},\uparrow}\hat{n}_{\vec{i},\downarrow}\rangle-n_{\vec{i}}^2=-\cG^e_{\vec{i}\vec{i}\vec{i}\vec{i}}.
\end{align}
Initially the total number of spin-up or spin-down electrons is
\begin{align}
  n_e&=\sum_{\vec{i}} n_{\vec{i}}=1,
\end{align}
and the system is uncorrelated, therefore the number of doublons is zero
\begin{align}
  n_d&=\sum_{\vec{i}} d_{\vec{i}}=0,
\end{align}
notice that uncorrelated part is subtracted in the definition of $d_{\vec{i}}$.

It is assumed that leads with $\gamma_\alpha=1$ and inverse temperature $\beta=100$ are attached suddenly at $t=0$,
i.\,e., $s_\alpha(t)=\theta(t)$, as the propagation starts at $t_i=0$. Since the central system and the leads are
initially not equilibrated, the relaxation gives rise to a small flow of charge in and out of the system at $t\simeq
0.5$. After this kink, $n_e$ remains constant, and the build up of $e$-$e$ and $e$-ph correlations takes place. This is
reflected in the rapid growth of the doublon population, primarily driven by pure electronic scattering, and its
oscillations resulting from interactions with phonons. Simultaneously, electrons and doublons propagate through the
lattice with distinct group velocities until they reach the boundaries. It is important to note that the travel times
for electrons and doublons are significantly different (compare the length of plateau in the plots of $n_e$ and
$n_d$). After reaching the boundaries, electrons tunnel to the leads indirectly reducing the number of doublons in the
system. The rate of tunneling strongly depends on the presence of phonons and on the level of approximation used. It can
be seen, for instance, that $T^{pp}$ approximation for the electron and Fan-Migdal (or $GD$) approximation for the
phonon self-energies, respectively, predicts the slowest decline of populations (smallest current). In the absence of
$e$-ph coupling ($g=0$) electrons leak to the leads faster, whereas the mean-field treatment of both interactions, which
does not capture the formation of bound doublon or polaron states, results in the fastest depletion of the central
region.
\section{Conclusions}
We presented a NEGF toolbox for the simulation of the coupled electron-boson dynamics in correlated materials including
the treatment of open systems with fermionic bath. The merits of our implementation in the \cheers{} package are as
follows: (1) all fundamental conservation laws are satisfied independently of the method~\cite{karlsson_fast_2021}; (2)
the ODE nature of the EOM allows one to address phenomena occurring at different time scales through a save-and-restart
procedure; and (3) as a byproduct of the calculations we have access to the spatially nonlocal correlators $\mcG^{e}(t)$
and $\mcG^{b}(t)$, as well as to the embedding correlator $\mathcal{G}^{\text{em}}$ that allows evaluation of currents
according to the Meir-Wingreen formula~\cite{tuovinen_time-linear_2022}. The ODE formulation naturally accommodates
parallel computations, adaptive time-stepping implementations, and restart protocols, thereby paving the way for
conducting first principles simulations of multiscale systems. As an illustration of our method, we investigated a 1D
Holstein-Hubbard model in a setup corresponding to the conduction band of a photovoltaic device. We found that the
formation of bound doublonic and polaronic states reduces the photo-current.
\section*{Supporting Information}
Supporting Information is available from the Wiley Online Library or from the authors. At present, the code is given
only for collaborative purposes.  The interested reader can contact the authors of the code.
\section*{Acknowledgements}
This research is part of the project No. 2021/43/P/ST3/03293 co-funded by the National Science Centre and the European
Union’s Horizon 2020 research and innovation programme under the Marie Sklodowska-Curie grant agreement no. 945339
(Y.P.).  G.S. and E.P. acknowledge funding from MIUR PRIN Grant No. 20173B72NB, from the INFN17-Nemesys project.
G.S. and E.P.  acknowledge Tor Vergata University for financial support through projects ULEXIEX and TESLA.


\begin{thebibliography}{10}
\providecommand{\url}[1]{\texttt{#1}}
\providecommand{\urlprefix}{URL }

\bibitem{basov_towards_2017}
D.~N. Basov, R.~D. Averitt, D.~Hsieh,
\newblock \emph{Nat. Mater.} \textbf{2017}, \emph{16}, 11 1077.

\bibitem{stefanucci_nonequilibrium_2013}
G.~Stefanucci, R.~van Leeuwen,
\newblock \emph{Nonequilibrium {Many}-{Body} {Theory} of {Quantum} {Systems}:
  {A} {Modern} {Introduction}},
\newblock Cambridge University Press, Cambridge, \textbf{2013}.

\bibitem{kwong_real-time_2000}
N.-H. Kwong, M.~Bonitz,
\newblock \emph{Phys. Rev. Lett.} \textbf{2000}, \emph{84}, 8 1768.

\bibitem{dahlen_solving_2007}
N.~E. Dahlen, R.~van Leeuwen,
\newblock \emph{Phys. Rev. Lett.} \textbf{2007}, \emph{98}, 15 153004.

\bibitem{myohanen_many-body_2008}
P.~My\"{o}h\"{a}nen, A.~Stan, G.~Stefanucci, R.~van Leeuwen,
\newblock \emph{Eurphys. Lett.} \textbf{2008}, \emph{84} 67001.

\bibitem{galperin_linear_2008}
M.~Galperin, S.~Tretiak,
\newblock \emph{J. Chem. Phys.} \textbf{2008}, \emph{128}, 12 124705.

\bibitem{myohanen_kadanoff-baym_2009}
P.~My\"{o}h\"{a}nen, A.~Stan, G.~Stefanucci, R.~van Leeuwen,
\newblock \emph{Phys. Rev. B} \textbf{2009}, \emph{80}, 11 115107.

\bibitem{von_friesen_successes_2009}
M.~P. von Friesen, C.~Verdozzi, C.-O. Almbladh,
\newblock \emph{Phys. Rev. Lett.} \textbf{2009}, \emph{103}, 17 176404.

\bibitem{bittner_coupled_2018}
N.~Bittner, D.~Gole\v{z}, H.~U.~R. Strand, M.~Eckstein, P.~Werner,
\newblock \emph{Phys. Rev. B} \textbf{2018}, \emph{97}, 23 235125.

\bibitem{schuler_time-dependent_2016}
M.~Sch\"{u}ler, J.~Berakdar, Y.~Pavlyukh,
\newblock \emph{Phys. Rev. B} \textbf{2016}, \emph{93}, 5 054303.

\bibitem{schuler_nessi_2020}
M.~Sch\"{u}ler, D.~Gole\v{z}, Y.~Murakami, N.~Bittner, A.~Herrmann, H.~U.
  Strand, P.~Werner, M.~Eckstein,
\newblock \emph{Comp. Phys. Commun.} \textbf{2020}, \emph{257} 107484.

\bibitem{kaye_low_2021}
J.~Kaye, D.~Golez,
\newblock \emph{SciPost Physics} \textbf{2021}, \emph{10}, 4 091.

\bibitem{meirinhos_adaptive_2022}
F.~Meirinhos, M.~Kajan, J.~Kroha, T.~Bode,
\newblock \emph{SciPost Physics Core} \textbf{2022}, \emph{5}, 2 030.

\bibitem{dong_excitations_2022}
X.~Dong, E.~Gull, H.~U.~R. Strand,
\newblock \emph{Phys. Rev. B} \textbf{2022}, \emph{106}, 12 125153.

\bibitem{lipavsky_generalized_1986}
P.~Lipavský, V.~\v{S}pi\v{c}ka, B.~Velický,
\newblock \emph{Phys. Rev. B} \textbf{1986}, \emph{34}, 10 6933.

\bibitem{bostrom_charge_2018}
E.~V.~n. Bostr\"{o}m, A.~Mikkelsen, C.~Verdozzi, E.~Perfetto, G.~Stefanucci,
\newblock \emph{Nano Lett.} \textbf{2018}, \emph{18}, 2 785.

\bibitem{covito_real-time_2018}
F.~Covito, E.~Perfetto, A.~Rubio, G.~Stefanucci,
\newblock \emph{Phys. Rev. A} \textbf{2018}, \emph{97}, 6 061401(R).

\bibitem{latini_charge_2014}
S.~Latini, E.~Perfetto, A.-M. Uimonen, R.~van Leeuwen, G.~Stefanucci,
\newblock \emph{Phys. Rev. B} \textbf{2014}, \emph{89}, 7 075306.

\bibitem{maansson_real-time_2021}
E.~P. M{\aa}nsson, S.~Latini, F.~Covito, V.~Wanie, M.~Galli, E.~Perfetto,
  G.~Stefanucci, H.~H{\"u}bener, U.~De~Giovannini, M.~C. Castrovilli, et~al.,
\newblock \emph{Commun. Chem.} \textbf{2021}, \emph{4}, 1 73.

\bibitem{perfetto_ultrafast_2018}
E.~Perfetto, D.~Sangalli, A.~Marini, G.~Stefanucci,
\newblock \emph{J. Phys. Chem. Lett.} \textbf{2018}, \emph{9}, 6 1353.

\bibitem{perfetto_first-principles_2019}
E.~Perfetto, D.~Sangalli, M.~Palummo, A.~Marini, G.~Stefanucci,
\newblock \emph{J. Chem. Theory Comput.} \textbf{2019}, \emph{15}, 8 4526.

\bibitem{perfetto_ultrafast_2020}
E.~Perfetto, A.~Trabattoni, F.~Calegari, M.~Nisoli, A.~Marini, G.~Stefanucci,
\newblock \emph{J. Phys. Chem. Lett.} \textbf{2020}, 9.

\bibitem{perfetto_first-principles_2015}
E.~Perfetto, A.-M. Uimonen, R.~van Leeuwen, G.~Stefanucci,
\newblock \emph{Phys. Rev. A} \textbf{2015}, \emph{92}, 3 033419.

\bibitem{perfetto_real-time_2022}
E.~Perfetto, Y.~Pavlyukh, G.~Stefanucci,
\newblock \emph{Phys. Rev. Lett.} \textbf{2022}, \emph{128}, 1 016801.

\bibitem{perfetto_real-time_2023}
E.~Perfetto, G.~Stefanucci,
\newblock \emph{Nano Lett.} \textbf{2023}, \emph{23}, 15 7029.

\bibitem{perfetto_cheers:_2018}
E.~Perfetto, G.~Stefanucci,
\newblock \emph{J. Phys. Condens. Matter} \textbf{2018}, \emph{30}, 46 465901.

\bibitem{schlunzen_achieving_2020}
N.~Schl\"{u}nzen, J.-P. Joost, M.~Bonitz,
\newblock \emph{Phys. Rev. Lett.} \textbf{2020}, \emph{124}, 7 076601.

\bibitem{joost_g1-g2_2020}
J.-P. Joost, N.~Schl\"{u}nzen, M.~Bonitz,
\newblock \emph{Phys. Rev. B} \textbf{2020}, \emph{101}, 24 245101.

\bibitem{pavlyukh_photoinduced_2021}
Y.~Pavlyukh, E.~Perfetto, G.~Stefanucci,
\newblock \emph{Phys. Rev. B} \textbf{2021}, \emph{104}, 3 035124.

\bibitem{karlsson_fast_2021}
D.~Karlsson, R.~van Leeuwen, Y.~Pavlyukh, E.~Perfetto, G.~Stefanucci,
\newblock \emph{Phys. Rev. Lett.} \textbf{2021}, \emph{127}, 3 036402.

\bibitem{tuovinen_time-linear_2022}
R.~Tuovinen, Y.~Pavlyukh, E.~Perfetto, G.~Stefanucci,
\newblock \emph{Phys. Rev. Lett.} \textbf{2023}, \emph{130} 246301.

\bibitem{meir_landauer_1992}
Y.~Meir, N.~S. Wingreen,
\newblock \emph{Phys. Rev. Lett.} \textbf{1992}, \emph{68}, 16 2512.

\bibitem{pavlyukh_time-linear_2022-1}
Y.~Pavlyukh, E.~Perfetto, D.~Karlsson, R.~van Leeuwen, G.~Stefanucci,
\newblock \emph{Phys. Rev. B} \textbf{2022}, \emph{105}, 12 125134.

\bibitem{pavlyukh_time-linear_2022}
Y.~Pavlyukh, E.~Perfetto, D.~Karlsson, R.~van Leeuwen, G.~Stefanucci,
\newblock \emph{Phys. Rev. B} \textbf{2022}, \emph{105}, 12 125135.

\bibitem{pavlyukh_electron_2013}
Y.~Pavlyukh, J.~Berakdar,
\newblock \emph{Comp. Phys. Commun.} \textbf{2013}, \emph{184} 387.

\bibitem{joost_dynamically_2022}
J.-P. Joost, N.~Schl\"{u}nzen, H.~Ohldag, M.~Bonitz, F.~Lackner,
  I.~B\v{r}ezinov\'{a},
\newblock \emph{Phys. Rev. B} \textbf{2022}, \emph{105}, 16 165155.

\bibitem{cole_applications_2016}
D.~J. Cole, N.~D.~M. Hine,
\newblock \emph{J. Phys. Condens. Matter} \textbf{2016}, \emph{28}, 39 393001.

\bibitem{weinkauf_highly_1996}
R.~Weinkauf, P.~Schanen, A.~Metsala, E.~Schlag, M.~B{\"u}rgle, H.~Kessler,
\newblock \emph{The Journal of Physical Chemistry} \textbf{1996}, \emph{100},
  47 18567.

\bibitem{kuleff_multielectron_2005}
A.~I. Kuleff, J.~Breidbach, L.~S. Cederbaum,
\newblock \emph{J. Chem. Phys.} \textbf{2005}, \emph{123}, 4 044111.

\bibitem{ayuso_ultrafast_2017}
D.~Ayuso, A.~Palacios, P.~Decleva, F.~Mart\'{i}n,
\newblock \emph{Phys. Chem. Chem. Phys.} \textbf{2017}, \emph{19}, 30 19767.

\bibitem{kim_pubchem_2021}
S.~Kim, J.~Chen, T.~Cheng, A.~Gindulyte, J.~He, S.~He, Q.~Li, B.~A. Shoemaker,
  P.~A. Thiessen, B.~Yu, L.~Zaslavsky, J.~Zhang, E.~E. Bolton,
\newblock \emph{Nucleic Acids Res.} \textbf{2021}, \emph{49}, D1 D1388.

\bibitem{pritchard_new_2019}
B.~P. Pritchard, D.~Altarawy, B.~Didier, T.~D. Gibson, T.~L. Windus,
\newblock \emph{J. Chem. Inf. Model.} \textbf{2019}, \emph{59}, 11 4814.

\bibitem{kloss_multiset_2019}
B.~Kloss, D.~R. Reichman, R.~Tempelaar,
\newblock \emph{Phys. Rev. Lett.} \textbf{2019}, \emph{123}, 12 126601.

\bibitem{bergamaschi_halogen_2018}
G.~Bergamaschi, L.~Lascialfari, A.~Pizzi, M.~I. Martinez~Espinoza, N.~Demitri,
  A.~Milani, A.~Gori, P.~Metrangolo,
\newblock \emph{Chem. Comm.} \textbf{2018}, \emph{54}, 76 10718.

\bibitem{murkakami_nonequilibrium_2017}
Y.~Murakami, N.~Tsuji, M.~Eckstein, P.~Werner,
\newblock \emph{Phys. Rev. B} \textbf{2017}, \emph{96} 045125.

\bibitem{hu_communication:_2010}
J.~Hu, R.-X. Xu, Y.~Yan,
\newblock \emph{J. Chem. Phys.} \textbf{2010}, \emph{133} 101106.

\bibitem{pavarini_dynamical_2014}
K.~Held,
\newblock In E.~Pavarini, E.~Koch, D.~Vollhardt, A.~I. Lichtenstein, editors,
  \emph{{DMFT} at 25: infinite dimensions: lecture notes of the {Autumn}
  {School} on {Correlated} {Electrons} 2014}. Forschungszentrum J\"{u}lich,
  Zentralbibliothek, Verl, J\"{u}lich, \textbf{2014}.

\bibitem{weber_electronic_2023}
M.~Weber, J.~K. Freericks,
\newblock \emph{Phys. Rev. Lett.} \textbf{2023}, \emph{130}, 26 266401.

\end{thebibliography}
\end{document}